\documentclass[a4paper]{jpconf}
\usepackage{graphicx}

\newcommand{\ttbar}{\ensuremath{t\overline{t}}}
\newcommand{\ljets}{\mbox{$\ell$+jets}}
\newcommand{\dil}{\mbox{$\ell\ell$}}
\newcommand{\met}    {\mbox{$\not\!\!E_T$}}

\newcommand{\ifb}{\mbox{fb$^{-1}$}}

\begin{document}
\title{Vector boson+jets as background to top-quark physics at Tevatron and LHC}

\author{Fr\'ed\'eric D\'eliot, {\rm \small on behalf of the ATLAS, CDF, CMS and D0 Collaborations}}

\address{CEA-Saclay, DSM/Irfu/SPP, bat 141, 91191 Gif sur Yvette Cedex, France}

\ead{frederic.deliot@cea.fr}

\begin{abstract}
I review here the latest measurements related to vector boson production in association with
light or heavy flavor jets at the Tevatron and at the LHC. 
The methods to estimate the $W$+jets and $Z$+jets background in top-quark analyses are also
presented.
\end{abstract}

\section{Introduction}
Vector boson production in association with jets ($V$+jets) is an important process.
First, it allows to test perturbative quantum chromodynamics (QCD) in a stringent way.
It is also used to check Monte Carlo (MC) simulation and stands as one of the favoured processes
to tune this simulation. 
These processes are particularly interesting because they allow precision QCD tests
with multiple renormalisation or factorisation scales set by the vector boson mass
and by the associated light or heavy-flavor jet transverse momentum.
Precise measurements of the $V$+jets properties are therefore of primary importance.

In top-quark physics, $V$+jets production is one of the main backgrounds in the lepton+jets 
(\ljets) and dilepton (\dil) channels. In both cases, vector boson production in association with
heavy quarks (mainly $b$-quarks) leads to the same final state as the production
of a top-antitop quark pair (\ttbar).
From a theoretical point of view, predictions exist for $V+\le 3$~jets at the Tevatron 
and $V+\le 4$~jets at the LHC.

Different jet reconstruction algorithms are used at the Tevatron and the LHC.
At the Tevatron, the midpoint cone algorithm is used with a cone radius of 
$\Delta R = \sqrt{\Delta \eta^2 + \Delta \Phi^2} = 0.4$ for CDF and $\Delta R = 0.5$ for D0.
The anti-kt algorithm is used at the LHC with $\Delta R = 0.4$ for ATLAS~\cite{atlas} and $\Delta R = 0.5$
for CMS~\cite{cms} which also exploits the technique of particle flow for jet reconstruction.
The LHC results presented in this article are based on 7~TeV data.
These reconstructed jets are calibrated for the electromagnetic and hadronic scales.
For the MC/data distributions shown in the following, the measurements are corrected 
at the particle level within the acceptance cuts. This means that the results are valid
for specific ranges of lepton and jets kinematics.

\section{$V$+jets experimental measurements}

\subsection{$W$+jets experimental measurements}

\begin{figure}[h]
\begin{center}
\begin{minipage}{11pc}
\includegraphics[width=11pc]{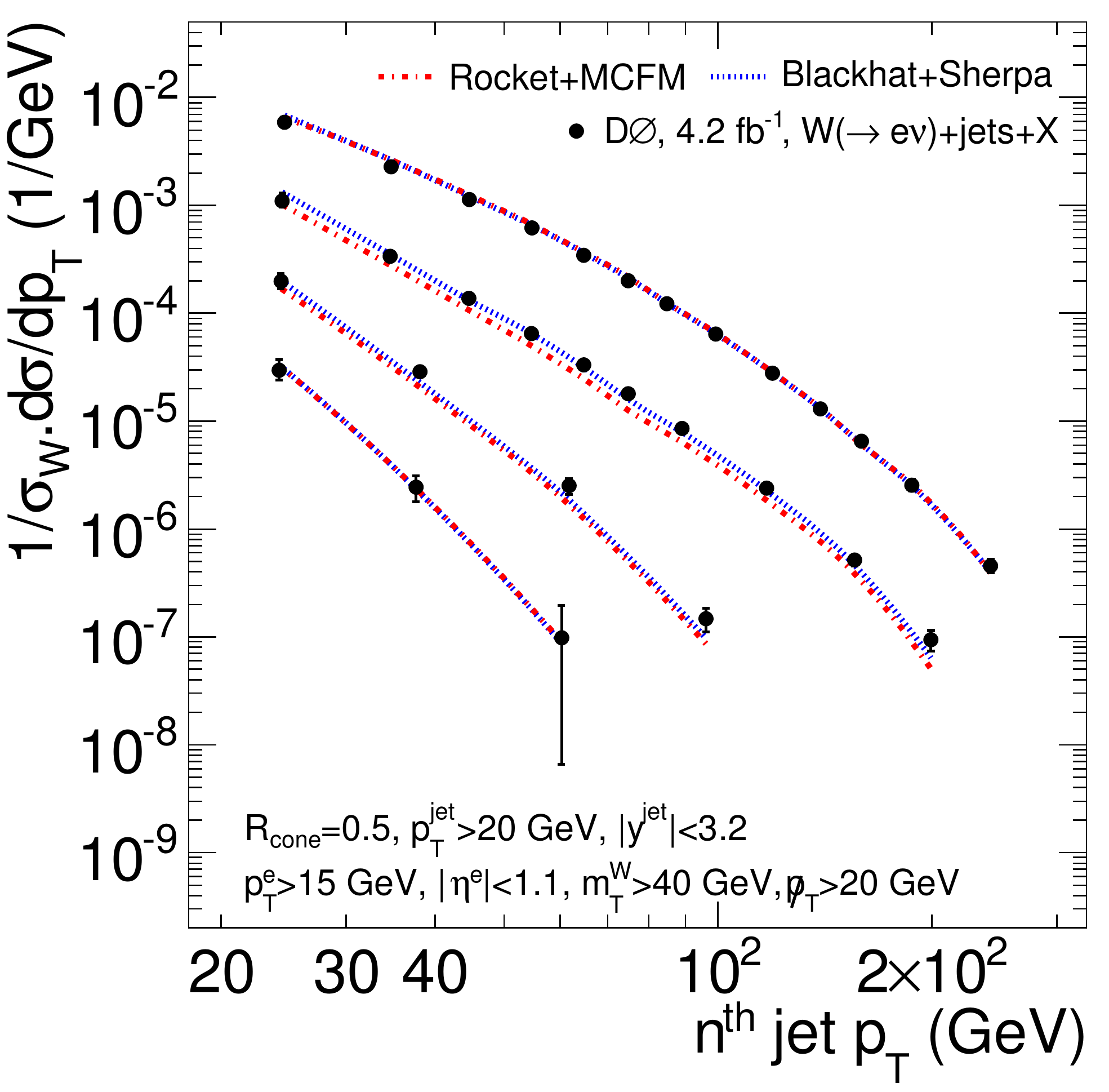}
\end{minipage}\hspace{0.5pc}%
\begin{minipage}{11pc}
\includegraphics[width=11pc]{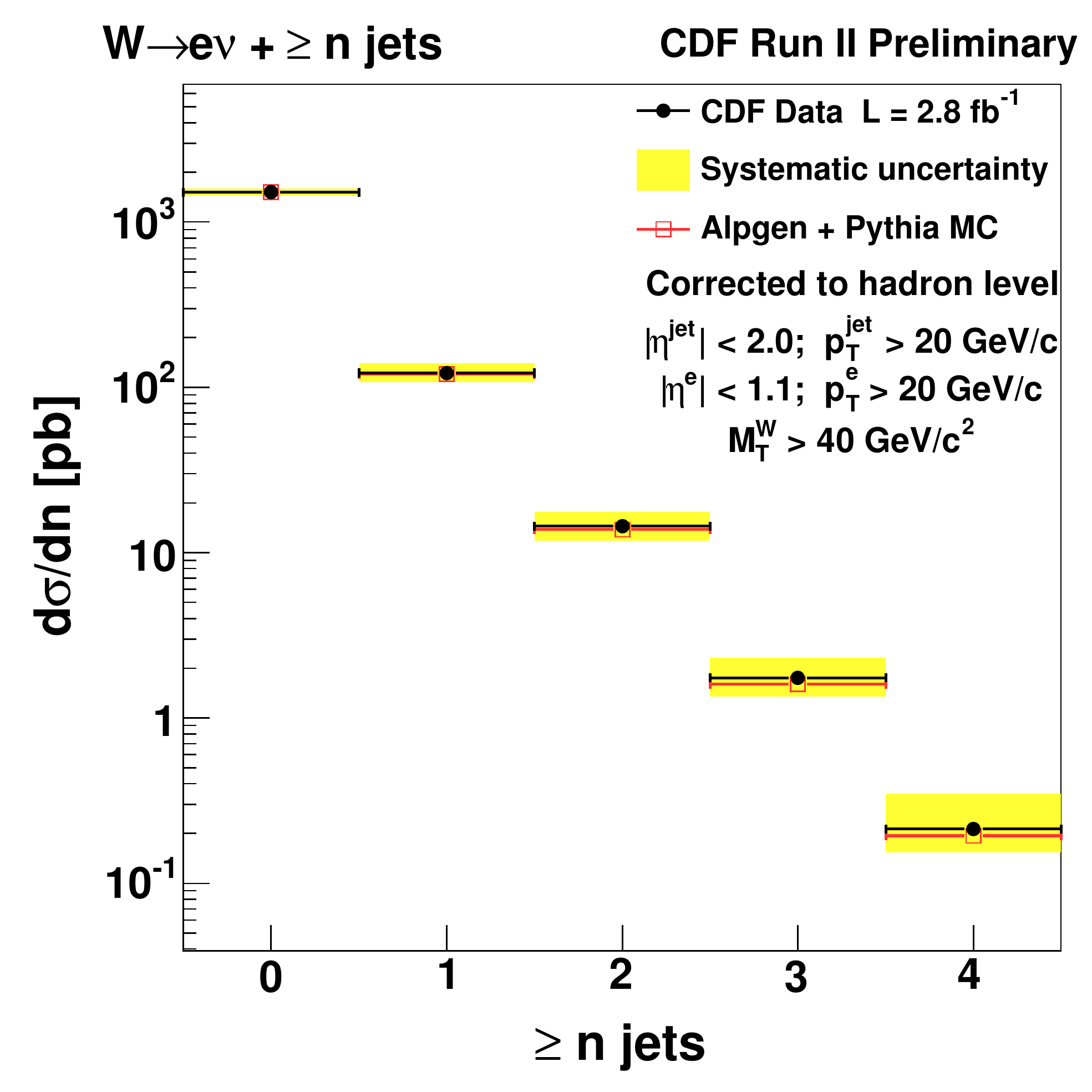}
\end{minipage} 
\caption{Measured W +n jet differential cross section at D0 for jet multiplicities $n =1-–4$~\cite{d0w} (left) 
and jet multiplicity 
measurement normalized to the total inclusive $W \to e \nu$ cross section at CDF~\cite{cdfw} (right).}
\label{fig:wjetstev}
\end{center}
\end{figure}

At the Tevatron the $W$+jets measurements rely on jets with
$p_T > 20$~GeV and $|\eta|<2$ for CDF or $|y|<3.2$ for D0. 
The measured differential cross sections are normalized to the total inclusive $W$ production cross
section so that cancellations of some systematic uncertainties occur. In some cases, the normalization
is rather performed using the next-to-next-leading order (NNLO) inclusive $W$ prediction.
The correction for reconstruction effects is completed by bin-by-bin unfolding in CDF or by regularized
matrix inversion in D0. The relative uncertainty on the results range from around 4\% 
$W+1$~jet to 20\% for $W+3$~jets. 
In general, comparisons with normalized multileg generators or next-to-leading order (NLO) simulations
show a good agreement. The predictions and measurements for the normalized differential cross section of
$W+e \nu \ge n$~jets in CDF~\cite{cdfw} and D0~\cite{d0w} are shown in Figure~\ref{fig:wjetstev}.

\begin{figure}[h]
\begin{center}
\begin{minipage}{11pc}
\includegraphics[width=11pc]{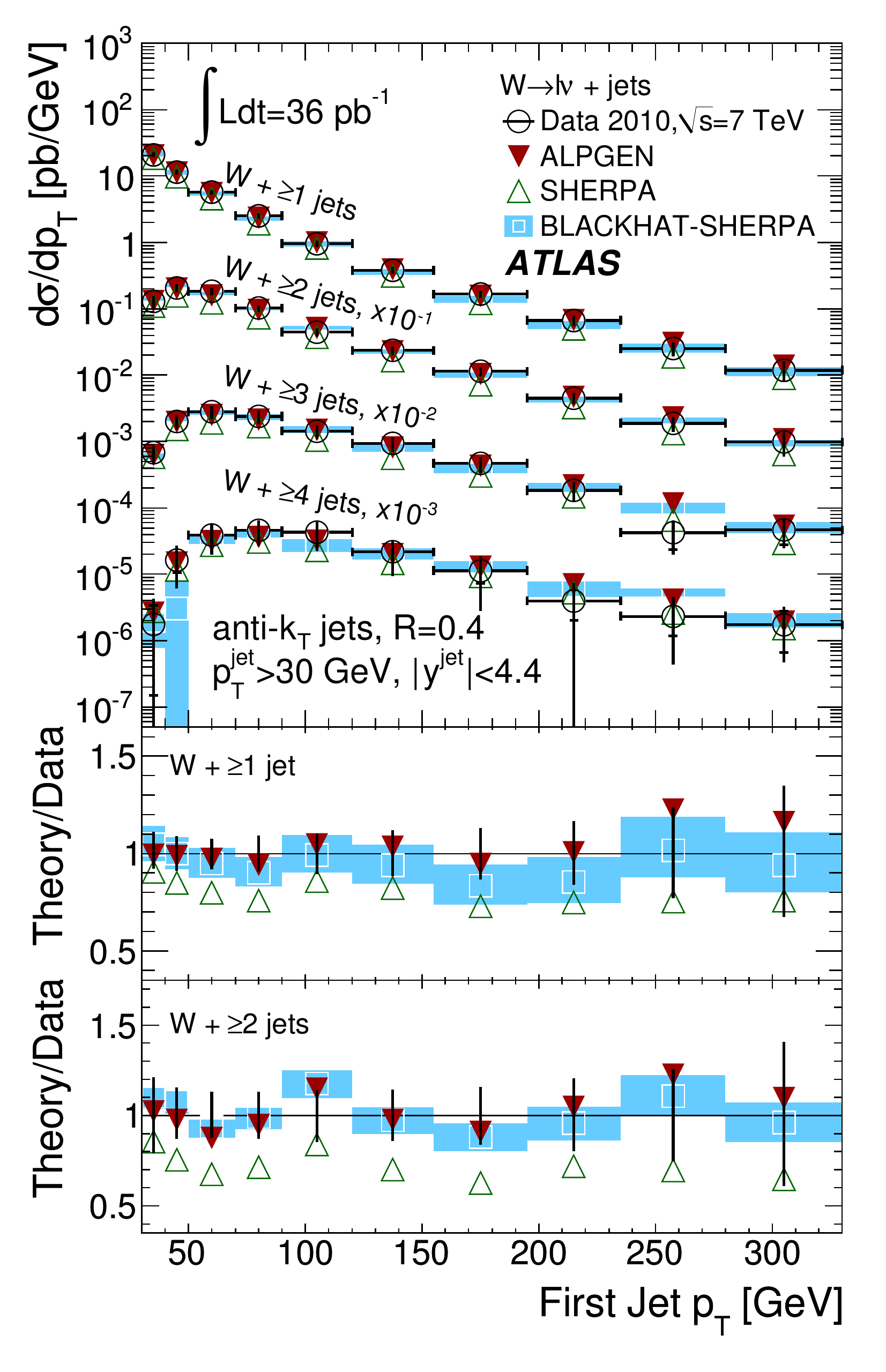}
\end{minipage}\hspace{0.5pc}%
\begin{minipage}{16pc}
\includegraphics[width=16pc]{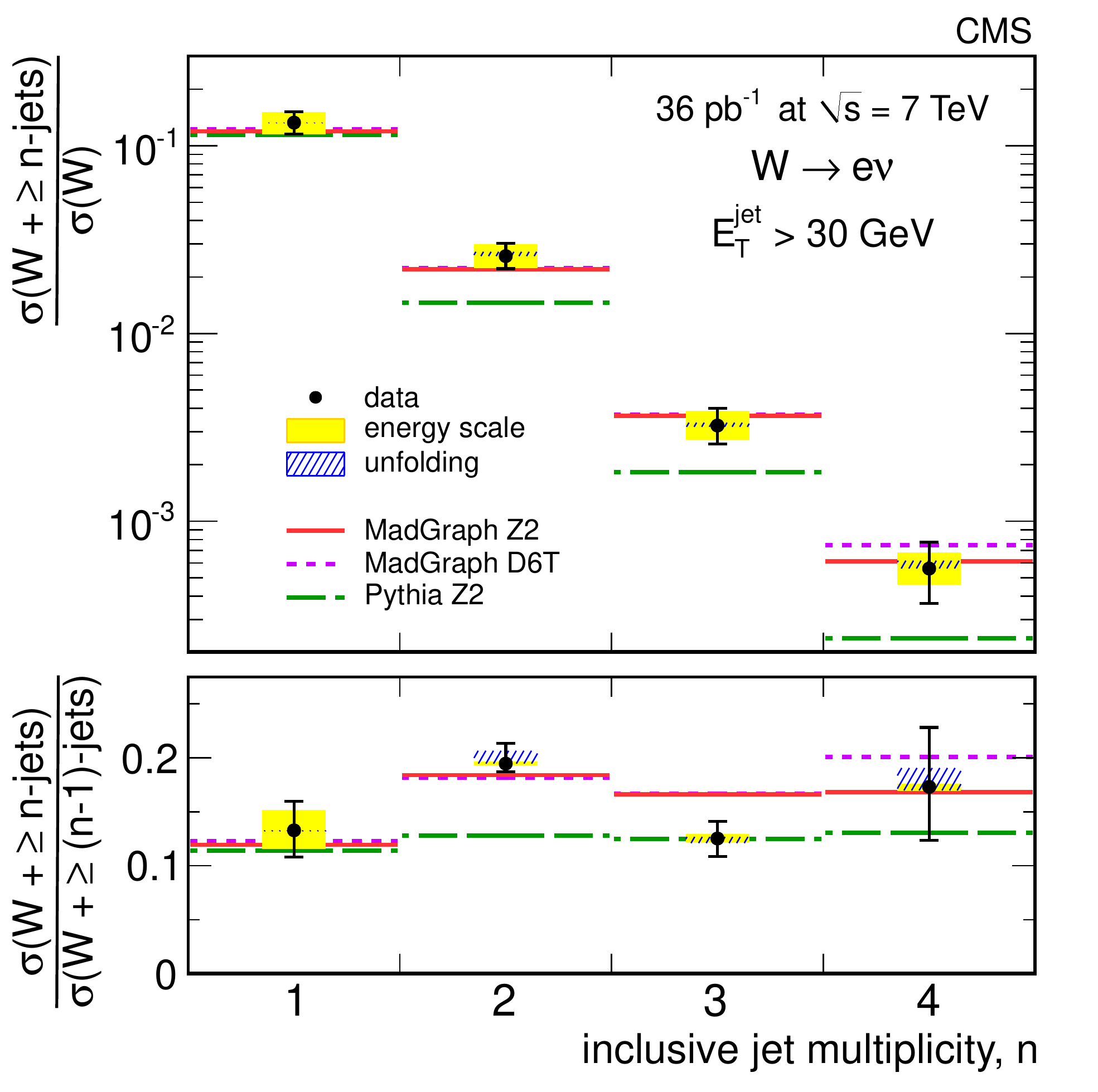}
\end{minipage} 
\caption{W+jets cross section as a function of the $p_T$ of the first jet in the event in ATLAS~\cite{atlasw} 
(left) and the ratio $\sigma(W+n \; jets)/\sigma(W)$ in the electron channel compared to the expectations 
in CMS~\cite{cmsw} (right).}
\label{fig:wjetslhc}
\end{center}
\end{figure}

At the LHC, jets with $p_T > 20$ or 30~GeV and $|y|<4.4$ for ATLAS or $|\eta|<2.4$ for CMS are selected.
The normalization for the differential cross section measurements is performed in the same way as at the Tevatron.
The correction for reconstruction effects is based on iterative Bayesian unfolding in ATLAS or singular
value decomposition (SVD) in CMS. The uncertainties for the measurements of the $W$ production vary 
from 8\% for the $W+1$~jet production to 20\% for $W+3$~jets~\cite{atlasw, cmsw}. 
At the LHC also, the comparison between multileg or
NLO generators shows good agreement.
Figure~\ref{fig:wjetslhc} shows the MC/data comparison for some differential cross section measurements
in ATLAS and CMS.

At the Tevatron, D0 and CDF also measure the inclusive $W+c$~jets cross sections.
In addition, CDF studies the $W+1,2b$~jets cross sections.
$W+c$~jets production is interesting since it is sensitive to the s~quark content of the proton
and the process is calculable in multi-scale perturbative QCD. 
MC predicts that the process $W+c$ represents around 5\% of the $W+1$~jet production (when the jet has
$p_T>10$~GeV). Soft lepton tagging is used to identify the $W+c$ production. 
With 4.3~\ifb\ of data, CDF measures $\sigma(W+c) \cdot Br(W \to \ell \nu) = 13.3^{+3.3}_{-2.9} {\rm (stat+syst)}$~pb
for jets with $p_T>20$~GeV and $|\eta|<1.5$ in agreement with the NLO prediction of $11.3 \pm 2.2$~pb.
With 1~\ifb\ of data, D0 measures 
$\frac{\sigma(W+c)}{\sigma(W+jets)} = 7.4 \pm 1.9 {\rm (stat)}^{+1.2}_{-1.4} {\rm (syst)}$\%~\cite{d0wc}
in agreement with the prediction from Alpgen+Pythia simulation of $4.4 \pm 0.3$\%.
With 1.9~\ifb\ of data, CDF also measures the cross section with at least one tight $b$-tagged jet 
using a template fit of the vertex mass
to get $\sigma(W+b/bb) \cdot Br(W \to \ell \nu) = 2.74 \pm 0.27 {\rm (stat)} \pm 0.42 {\rm (syst)}$~pb~\cite{cdfwb}
that appears to be higher by around $3 \sigma$ than the NLO prediction of $1.22 \pm 0.14 {\rm (syst)}$~pb.
The LHC experiments also study the $W+$heavy flavor jet production.
Using events with one $b$-tag and a template fit to the vertex mass distribution in 35~pb$^{-1}$, ATLAS measures 
$\sigma(W+b/bb) \cdot Br(W \to \ell \nu) = 10.2 \pm 1.9 {\rm (stat)} \pm 2.6 {\rm (syst)}$~pb within the 
fiducial cuts~\cite{atlaswb} which is $1.8 \sigma$ higher than the NLO prediction of $4.8 \pm 1.0$~pb.
Using events with one tag and a template fit of a secondary vertex discriminant in 36~pb$^{-1}$, CMS measures
$\frac{\sigma(W+c)}{\sigma(W+jets)} = 14.3 \pm 1.5 {\rm (stat)} \pm 2.4 {\rm (syst)}$\% using jets with
$p_T>20$~GeV and $|\eta|<2.1$~\cite{cmswc} in agreement with the NLO prediction of $13 \pm 2$\%.

\subsection{$Z$+jets experimental measurements}

Studies of $Z$+jets production at the Tevatron use jets with $p_T>20$ or 30~GeV and $|\eta|<2.1$ in CDF and
$|y|<2.5$ at D0. 
The uncertainties of the differential measurements typically range from 7 to 10\% (for $Z+2$~jets).
Measurements~\cite{d0z} are compared to leading order (LO), multileg and NLO generators.
A good agreement is reached with multileg and NLO generators. The measured differential cross sections as a function
of jet multiplicity or jet momentum are generally more precise than MC predictions.

Measurements are also performed at the LHC using 36~pb$^{-1}$ of data using jets with $p_T>30$~GeV and
$|y|<4.4$ for ATLAS and $|\eta|<2.4$ for CMS. For reconstruction corrections, ATLAS uses a bin-by-bin unfolding
while CMS uses SVD. CMS also tests the so-called Berends-Giele scaling by measuring 
$\frac{\sigma(Z+n~jets)}{\sigma(Z+n+1~jets)} = \alpha + n \beta$ showing that $\beta$ is compatible with 0.
The uncertainties range between 13 to 20\%.
A good agreement is observed between data~\cite{atlasz, cmsw} and multileg and NLO generators.

At the Tevatron, CDF measures the $Z+b$~jets cross section using jets reconstructed with the midpoint cone algorithm
with $R=0.7$ and $p_T>20$~GeV and $|\eta|<1.5$ while the leptons are selected using a neural network.
Using 9.1~\ifb\ of data, the cross section is extracted using a template fit of the secondary vertex mass leading to:
$\frac{\sigma(Z+b~jets)}{\sigma(Z+jets)} = 2.08 \pm 0.18 {\rm (stat)} \pm 0.27 {\rm (syst)}$\% in agreement with
the predictions from MCFM and corresponding to a K-factor of 1.6 compared to the Alpgen prediction.
At D0, jets with $p_T>15$ or 20~GeV and $|\eta|<2.5$ are used. A template fit of a discriminant variable 
built with the secondary vertex mass and the probability from tracks to originate from the primary vertex
is used.
With 4.2~\ifb\ of data, D0 measures 
$\frac{\sigma(Z+b~jets)}{\sigma(Z+jets)} = 1.93 \pm 0.22 {\rm (stat)} \pm 0.15 {\rm (syst)}$\%~\cite{d0zb}
in agreement with the prediction from MCFM of $1.92 \pm 0.22$\%.

At the LHC, the $Z$+b~jets cross section is measured with jets with $p_T>25$~GeV and $|y|<2.1$ with a template 
fit of the secondary vertex mass 
yielding $3.55^{+0.82}_{-0.74} {\rm (stat)} ^{+0.73}_{-0.55} {\rm (syst)} \pm 0.12 {\rm (lumi)}$~pb~\cite{atlaszb}
in agreement with the predictions from MCFM of $3.88 \pm 0.58$~pb. 
With 2.2~\ifb\ of data, CMS measures independently the $Z+1b$ and $Z+2b$~jets cross sections using jets with $p_T>25$~GeV
and $|y|<2.1$ and a template fit to the secondary vertex mass of the leading jet.
The results give $\sigma(Z+b) = 5.84 \pm 0.08 {\rm (stat)} \pm 0.72 {\rm (syst)} ^{+0.25}_{-0.55} {\rm (theory)}$~pb~\cite{cmszb}
compatible with the prediction from MCFM, and 
$\sigma(Z+bb) = 0.37 \pm 0.02 {\rm (stat)} \pm 0.07 {\rm (syst)} \pm 0.02 {\rm (theory)}$~pb~\cite{cmszbb}
in agreement with the predictions from Madgraph.

\subsection{Summary of the $V$+jets measurements}

Even if a lot of $V$+jets measurements have been performed at the Tevatron and the LHC,
only a few of them have been listed above.
The $V$+light jets measurements agree with predictions from NLO and multileg generators (scaled
to the NNLO inclusive prediction). The measurements are already enabling to constrain the MC predictions
since their precisions are around 20\% for $W+3$~jets and 15\% for $Z+2$~jets.
In general, measurements of $V$+heavy flavor jets agree with predictions from NLO and scaled
multileg generators. However, some discrepancy exists in the measurement of the $W+b$~jets cross section
at CDF while the LHC measurements so far agree with the predictions for the same process.
The measurement uncertainties are generally larger than the ones on MC predictions.
Typically, the precisions obtained in data are around 25\% for $W+c$~jets, 20-30\% for $W+b$~jets
and around 15-30\% for $Z+b$~jets.
Further measurements with higher statistics are expected, both, from Tevatron and LHC in the near future.
LHC measurements with $\sqrt{s}=8$~TeV are also expected.

\section{$V$+jets background determination in top-quark analyses}

Depending on the experiments, on the colliders and on the analyses, 
different techniques are used to evaluate the background from $V$+jets in top-quark measurements.
Only some of them will be discussed below which are based on the methods used for \ttbar\ and single 
top-quark cross section measurements.
The general rule (which suffers from exception) is that the $V$+jet background shapes 
in top-quark analyses are usually taken from MC (either from Alpgen or Madgraph) while the 
$V$+jets normalization is usually measured directly on data. The ratio of $V$+heavy flavor (hf) jets
over $V$+light jets is rescaled using NLO predictions and/or checked or constrained using data itself.
When both light and hf jet MC samples are produced separately, there is some
overlap of hf jets in the light flavor jet samples because hf jets 
can originate from matrix element (ME) or parton shower (PS). A procedure to remove
the overlap needs to be established. A possible removal can be based on jet angles, i.e.
for a given heavy flavor jet, the jet from PS is kept if there is a jet $j$ with 
$\Delta R(j,hf)<0.4$ otherwise the jet from ME is kept.

\subsection{$W$+jets background determination}
The $W$+jets background is of particular importance in the lepton+jets channel.

At CDF, Alpgen+Pythia is used to model the shapes of the $W$+jets background.
In the \ttbar\ lepton+jets cross section measurements, the $W$+jets background is normalized 
to data before $b$-tagging and after subtraction of the other backgrounds.
The \ttbar\ cross section was measured using a fit to the distribution of a 
neural network based flavor separator in 9 samples ($1- \ge 5 $~jets, 1 or 2~$b$-tag)~\cite{cdfttbarw}. 
The background is constrained with the 1 or 2~jets 
sample with floating normalization for the $b\bar{b}$, $c\bar{c}$ and single $c$ MC samples.
Fitting the normalization allows to reduce the $W$-jets related-systematic uncertainties.
The measured K-factor related to Alpgen+Pythia is $K_{b\bar{b}} = 1.39^{+0.28}_{-0.22}$ 
in agreement with the theoretical prediction.
In CDF single top cross section measurements, the $W$+n~jets pretag data sample is 
used to scale the prediction from Alpgen.
The flavor separator is fitted in the 1~jet sample to get $K_{hf}=1.4 \pm 0.4$.
A 30\% uncertainty is assessed for the extrapolation to the $W$+2,3~jets samples.

D0 is also using Alpgen+Pythia for modelling the $W$+jets background.
In the \ttbar\ lepton+jets cross section measurement, the normalization is fitted after subtraction 
of the other background together with the \ttbar\ cross section in bins of number of 
jets. After checking on data, the NLO predictions for the ratio 
$R_{hf} = N(W+bb/cc)/N(W+light)=1.47 \pm 0.22$ 
and $R_c = N(W+c)/N(W+light) = 1.27 \pm 0.15$ 
are used which translates into an uncertainty of 0.2~pb on the measured \ttbar\ cross 
section~\cite{d0ttbarw}.
$R_{hf}$ can also be fitted together with the \ttbar\ cross section using samples with 1 or 2~jets,
with or without $b$-tagging leading to 
$R_{hf}= 1.55 \pm 0.09 ({\rm stat}) ^{+0.17}_{-0.19} ({\rm syst})$. The uncertainty on $R_{hf}$
translates into an uncertainty of 0.02~pb in the fitted \ttbar\ cross section.
In the single top quark cross section analysis, the $W$+jets background is normalized 
with the multijet background before $b$-tagging using the lepton $p_T$, the missing transverse
energy (\met) and the $W$ transverse mass distributions.
The $W$+hf scale factor is taken from the NLO predictions and cross-checked with the 0-tag
2 jet sample. A 12\% uncertainty is assigned to this normalization.
Figure~\ref{fig:ttbarwtev} shows the CDF and D0 distributions on control samples in the 
\ttbar\ cross section analyses.

\begin{figure}[h]
\begin{center}
\begin{minipage}{13pc}
\includegraphics[width=13pc]{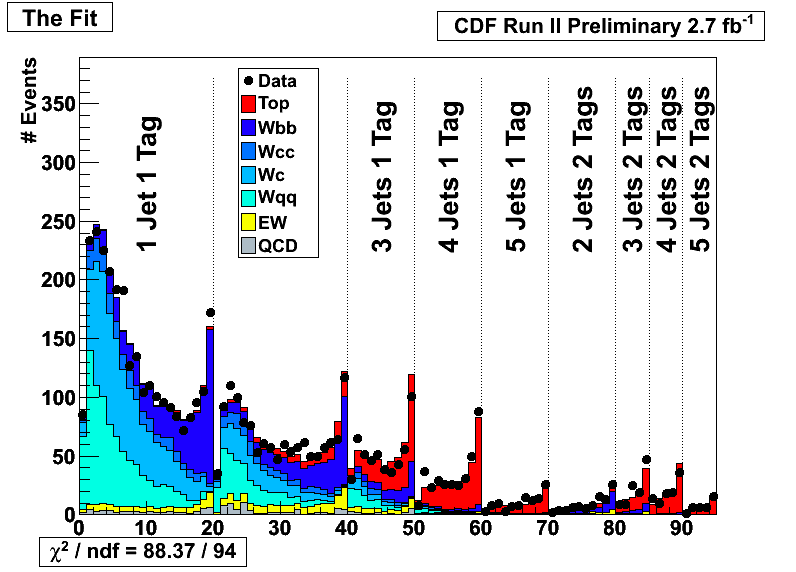}
\end{minipage}\hspace{0.5pc}%
\begin{minipage}{14pc}
\includegraphics[width=14pc]{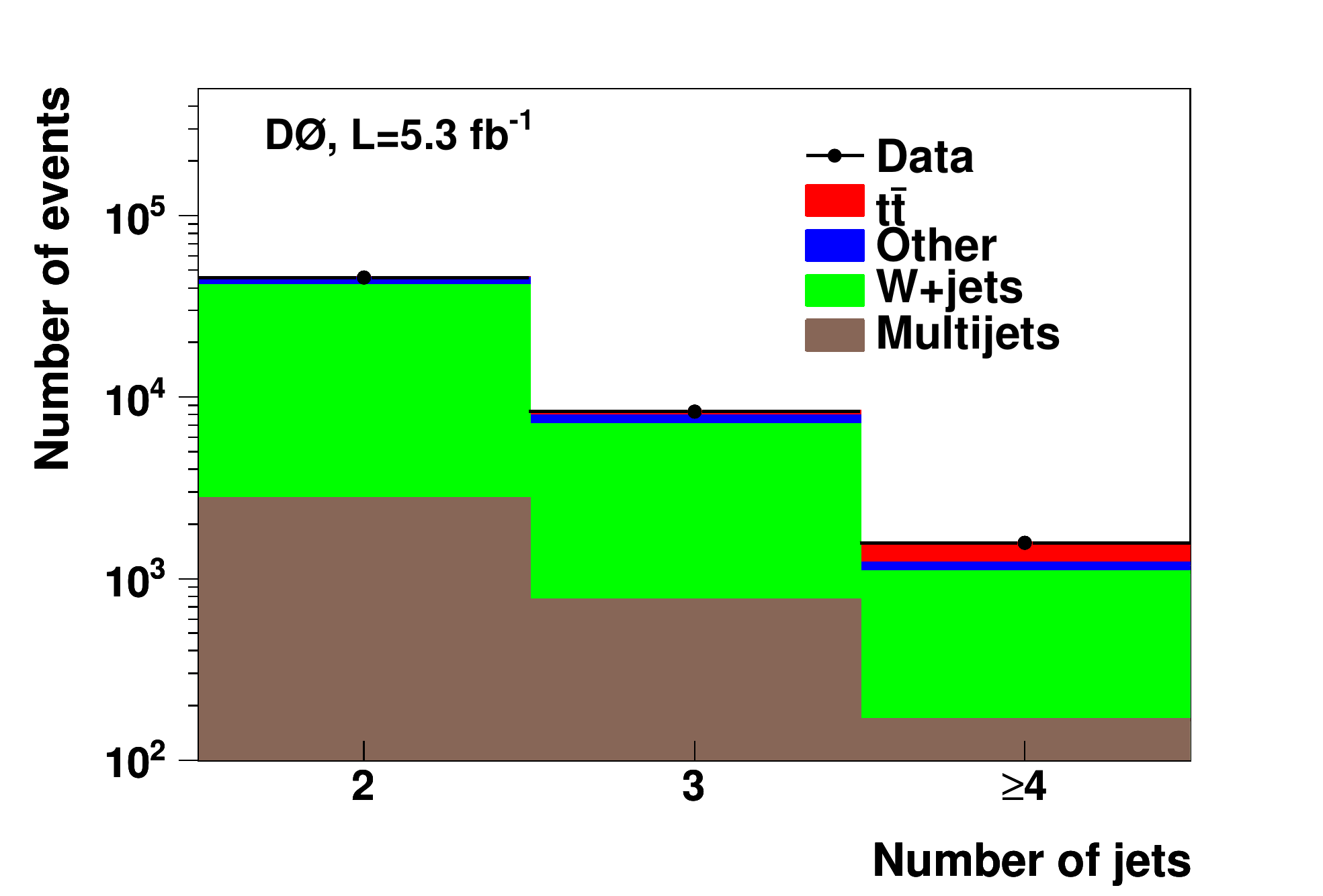}
\end{minipage}
\caption{Output of the flavor separator after the fit split by the number of jets and tags
in CDF~\cite{cdfttbarw} (left), jet multiplicity distribution for events with 0 $b$-tagged 
jets at D0~\cite{d0ttbarw} (right).}
\label{fig:ttbarwtev}
\end{center}
\end{figure}

In ATLAS, Alpgen+Herwig is used to generate the $W$+jets MC samples.
In the \ttbar\ lepton+jets cross section analysis, the $W$+jets normalization is fitted together
with the \ttbar\ cross section. Alternatively, the ratio of the number of produced
$W^+$ and $W^-$ is used. At the LHC, due to the quark content of the protons, more
$W^+$ are produced than $W^-$ in $W$+jets processes. 
The ratio $r_{MC} = W^+/W^-$ is well known theoretically.
Hence the total number of $W$+jets before tagging (in the 4 jet inclusive bin) can be 
extracted from the value of $r_{MC}$ in the MC with the formula: $N_{\ge 4 pretag} = \frac{r_{MC}+1}{r_{MC}-1}(D^+-D^-)$
where $D^+,D^-$ is the number of events with $W^+$ or $W^-$ in data after selection.
The hf scale factor is determined in the 1, 2 jet sample without and with $b$-tagging.
In single top cross section measurements, the $W$+jets sample is scaled by a factor K=1.2 
which corresponds to the inclusive NNLO predictions from FEWZ. The $W$+hf normalization is fitted
together with the signal.

CMS is using Madgraph+Pythia to generate the $W$+jets MC samples.
In their untagged \ttbar\ cross section analysis, a template fit of discriminating variables
(like the distribution of \met) allows to extract the $W$+jets
normalization and the \ttbar\ cross section. In the tagged analysis, a template fit of 
the secondary vertex mass is performed to get the $W$+jets normalization and the hf scale factor.
The extracted hf scale $K_{hf}$ is found to be slightly higher than the
NNLO predictions: $K_{W+b~jets} = 1.9^{+0.6}_{-0.5}$ and $K_{W+c~jets} = 1.4 \pm 0.2$.
For the single top quark cross section measurements, a template fit of the 
non-tagged jet pseudo-rapidities is exploited in sidebands using the 2 jets, 1 $b$-tag sample.
Figure~\ref{fig:ttbarwlhc} shows typical distributions for the $W$+jets, \ttbar\ 
combined fit at the LHC. 

\begin{figure}[h]
\begin{center}
\begin{minipage}{15pc}
\includegraphics[width=15pc]{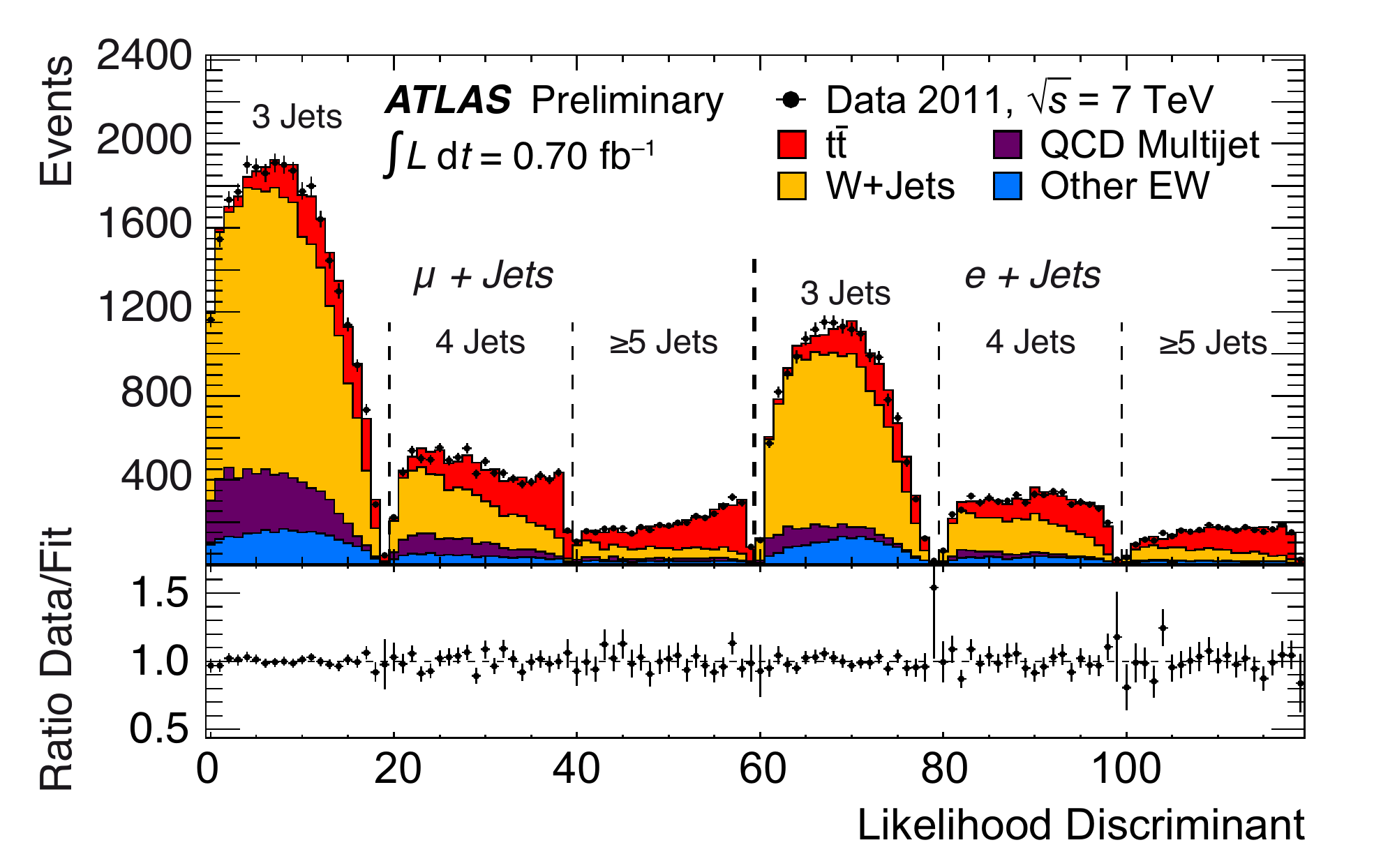}
\end{minipage}\hspace{0.5pc}%
\begin{minipage}{14pc}
\includegraphics[width=14pc]{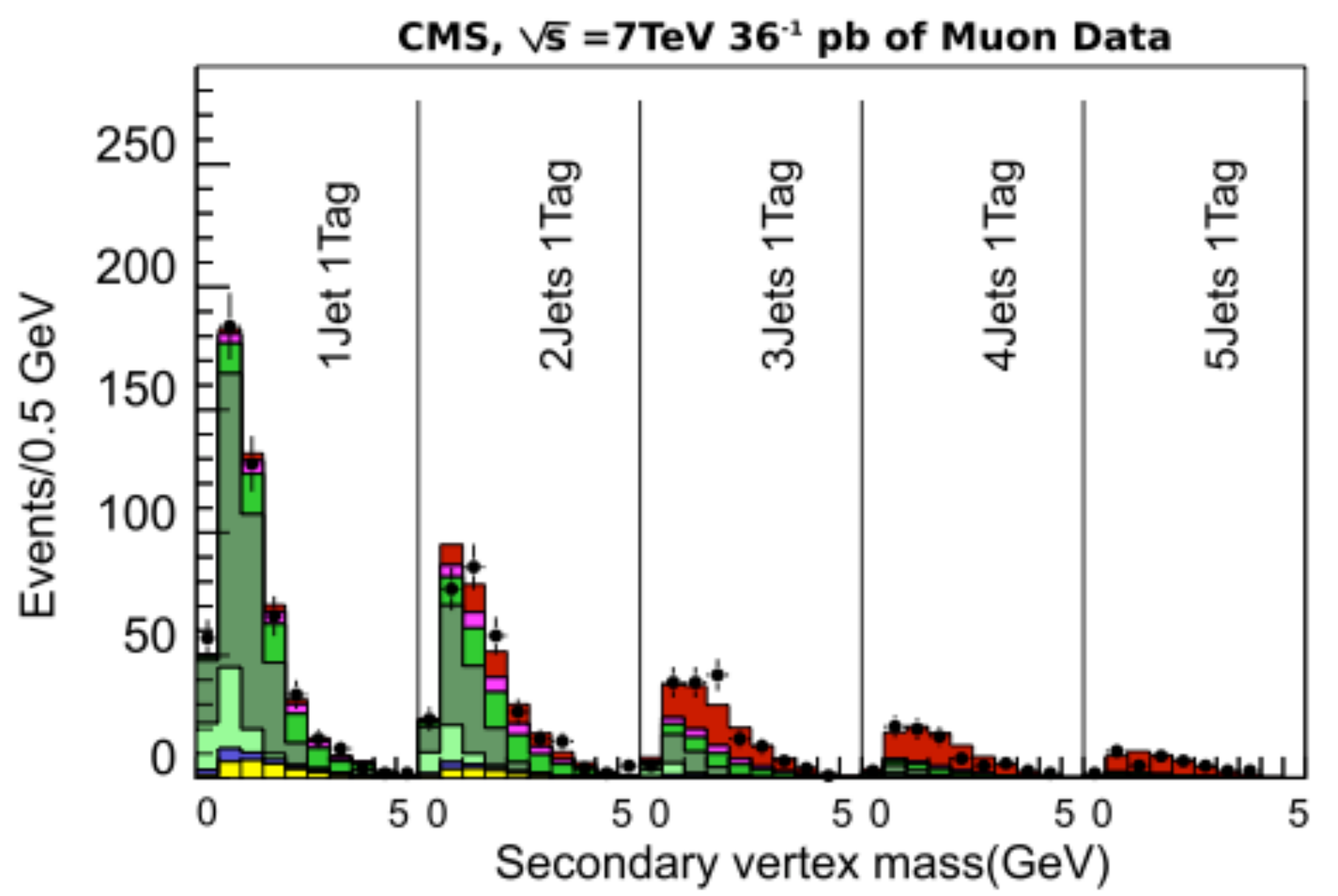}
\end{minipage}
\caption{Result of combined fit to data in the 3 to 5-jet bin in ATLAS~\cite{atlasttbarw} (left), 
Results of the combined fit in the muon channel for exactly 1 $b$-tag at CMS~\cite{cmsttbarw} 
(right).}
\label{fig:ttbarwlhc}
\end{center}
\end{figure}

\subsection{$Z$+jets background determination}
The $Z$+jets background is mostly relevant for the analyses in the dilepton channel.
As for the $W$+jets samples, the Tevatron experiments are using Alpgen+Pythia to generate
the $Z$+jets MC background while ATLAS is using Alpgen+Herwig and CMS Madgraph+Pythia.
Depending on the collaborations, the number of $Z$+light jets or $Z$+hf jets is taken
directly from the predictions of the multileg generators scaled to the NNLO computations, or
it is evaluated using a data-driven method.
The data-driven method scales the data in the Z mass window after \met\ selection.
The MC is used afterwards to predict the ratio of events inside and outside the Z mass region.
The data driven extraction agrees with the MC predictions but usually leads to smaller 
uncertainties (1-3\% uncertainties on the \ttbar\ cross section).

\section{Conclusion}
A lot of $V$+jets measurements have been performed by the CDF, D0, ATLAS and CMS experiments at 
the Tevatron and the LHC.
The measurements agree with the NLO MC predictions.
Multilegs generators are suitable to reproduce most of the $V$+jets kinematics.
Still, more results and checks are expected to come.
Various methods are used to measure the $V$+jets background in top-quark analyses.
Depending on the needed precision, multileg generators are scaled to NLO predictions
or data driven normalization are used or fit together with the top-quark signal.
Thanks to the advanced background estimation technique, the achieved
uncertainty on the $V$+jets background is usually found to be small.

\end{document}